\documentclass[pdftex,twocolumn,epjc3]{svjour3}          
\RequirePackage[utf8]{inputenc}
\RequirePackage[T1]{fontenc}

\smartqed  

\RequirePackage{graphicx}
\RequirePackage{mathptmx}      
\RequirePackage{flushend}
\RequirePackage[numbers,sort&compress]{natbib}
\RequirePackage[colorlinks,citecolor=blue,urlcolor=blue,linkcolor=blue]{hyperref}
\usepackage{amsfonts}
\usepackage{amssymb}

\journalname{Eur. Phys. J. C}

\begin{document}

\title{Jeans Instability in Non-Minimal Matter-Curvature Coupling Gravity}


\author{Cláudio Gomes\thanksref{e1,addr1}
}

\thankstext{e1}{e-mail: claudio.gomes@fc.up.pt}

\institute{Centro de Física do Porto, Rua do Campo Alegre s/n, 4169-007 Porto, Portugal\label{addr1}
}

\date{Received: date / Accepted: date}

\maketitle

\begin{abstract}
The weak field limit of the nonminimally coupled Boltzmann equation is studied, and relations between the invariant Bardeen scalar potentials are derived. The Jean's criterion for instabilities is found through the modified dispersion relation. Special cases are scrutinised and considerations on the model parameters are discussed for Bok globules.
\end{abstract}

\section{Introduction}

The Boltzmann equation is a fundamental description of the microscopic world and is derived from the Liouville equation in phase space considering collisions between particles. From the former, one can derive macroscopic equations, such as the Navier-Stokes equation for fluids and the virial theorem for gravitationally bound systems \cite{binney}, the \break Maxwell-Vlasov equations which characterise plasmas \cite{nicholson}, the quantum Bloch-Boltzmann equations for electrons \cite{pottier}, and the evolution of primordial elements' abundances in a de Sitter Universe \cite{turner}. From the Boltzmann equation it is also possible to build physical quantities from its moments, such as the particle number flux, the energy-momentum tensor or the entropy vector flux.

The Boltzmann equation is sensible to relativistic and quantum effects. In particular, it can be generalised in order to account for modified gravity models. In fact, despite its successful agreement with a vast plethora of observational data tests \cite{will,obgr}, General Relativity (GR) lacks a fully consistent quantum version of it and requires two dark components to match observations at astrophysical and cosmological scales, namely dark matter and dark energy, which have not been directly observed so far. Thus, several alternative theories of gravity have been proposed over the years in the literature. One of the simplest generalisations of GR is the so-called f(R) theories which replace the Ricci scalar by a generic function of it in the action functional (see  Refs.  \cite{fr1,fr2}  for review on this and Ref. \cite{extended} for a review on basic principles a gravity theory must obey and some extended theories of gravity). In fact, one specific proposal of such theories was firstly advanced in order to tackle the initial conditions problems of the standard Hot  Big  Bang  model, namely through a nonsingular isotropic homogeneous solution which accounted for inflation \cite{starobinsky}. Moreover, this model is still in excellent agreement with the most recent data from Planck mission \cite{planck}. (We refer the reader to Ref. \cite{exoticinflation} for a review of some exotic inflationary models in light of Planck data).

Furthermore, f(R) theories of gravity have been used to address the problems of dark matter and dark energy (see e.g. Refs. \cite{frdark1,frdark2}). It has also been found that by requiring that f(R) models to be regular at $R=0$ leads to a behaviour compatible with an effective cosmological constant in a sufficiently curved spacetime which disappears in flat spacetime \cite{starobinsky2}.

Another successful alternative theory of gravity in shedding some light in the above mentioned problems relies in an extension of f(R) theories with a non-minimal coupling between matter and curvature \cite{nmc}. In fact, it allows for a mimicking effect of dark matter effects at galaxies \cite{dm1} and clusters of galaxies scales \cite{dm2}, it has some bearings on the late time acceleration \cite{de1}, and is compatible with Planck's inflation data \cite{nmcinflation}, gravitational waves measurements \cite{nmcgw} and the modified virial theorem from the spherically relaxed Abell 586 cluster \cite{linmc}. 

In the weak field regime, this model yields a correction to the Newtonian potential \cite{martins}, and presents shock waves in the gravitational collapse \cite{numerico1,numerico2}. It has been recently applied to the generalisation of the Boltzmann equation for such scenario and its main implications were analysed in Ref. \cite{nmcboltzmann}. Therefore, it is important to study how this alternative model of gravity modifies the Jeans criterion for instability, which is responsible for the collapse of a gravitationally bound system, such as interstellar gas clouds whose internal pressure does not overcome gravity, ultimately leading to star formation.

This work is organised as follows: in section \ref{sec:nmc}, the non-minimal mater-curvature coupling model is introduced; in section \ref{sec:weakfield} we derive the weak field of the nonminimally coupled Boltzmann equation of Ref. \cite{nmcboltzmann}. In the following section, \ref{sec:jeans}, the Jean's criterion for instabilities is analysed and the implications for the parameters of the alternative gravity model are found. As an example, we test the viability of the model for Bok globules in Sec. \ref{sec:bok}. The conclusions are presented in Sec. \ref{sec:conclusions}.

\section{The non-minimal matter-curvature coupling model}
\label{sec:nmc}

The non-minimal matter-curvature coupling alternative gravity model (NMC) is defined from its action functional \cite{nmc}:
\begin{equation}
\label{modelo}
S=\int d^4x \sqrt{-g} \left[\kappa f_1\left(R\right) + f_2 \left(R\right)\mathcal{L}\right]~,
\end{equation}
where $f_1(R), f_2(R)$ are arbitrary functions of the curvature scalar $R$, $\kappa=c^4/(16\pi G)$ and $\mathcal{L}$ is the Lagrangian density of matter fields.

The metric field equations can be straightforwardly found by varying the previous action   with respect to the metric, $g_ {\mu\nu}$:
\begin{equation}
\label{fieldequations}
\Theta G_{\mu\nu} = \frac{1}{2\kappa}f_2(R) T_{\mu\nu} + \Delta_{\mu\nu}\Theta + \frac{1}{2}g_{\mu\nu}\left[f_1(R)-R\Theta\right] ~,
\end{equation}
where $\Theta:=\left( f'_1(R) + \frac{f'_2(R)\mathcal{L}}{\kappa}\right)$, $G_{\mu\nu}$ is the Einstein tensor, the primes denotes derivatives with respect to the curvature scalar, $f'_i(R) \equiv df_i(R)/dR$, and $\Delta_{\mu\nu}\equiv \nabla_{\mu}\nabla_{\nu} - g_{\mu\nu} \Box $. Moreover, General Relativity can be retrieved as the particular case when $f_1(R)=R$ and $f_2(R)=1$.

The trace of the metric field equations reads:
\begin{equation}
\label{trace}
\Theta R-2f_1(R)=\frac{1}{2\kappa}f_2(R)T -3\square \Theta  ~.
\end{equation}

In fact, one of the most striking features of this model consists in the covariant non-conservation of the energy-momentum tensor:
\begin{equation}
\label{nonconservation}
\nabla_{\mu}T^{\mu\nu} =  \left(g^{\mu\nu}\mathcal{L}-T^{\mu\nu}\right) \nabla_{\mu}\ln f_2(R) ~.
\end{equation}

This implies that for a perfect fluid, with $T_{\mu\nu}=(\rho+p)u^{\mu}u^{\nu}-p~g_{\mu\nu}$, test particles do not follow geodesic lines given the presence of an extra force term in the geodesics equation \cite{nmc}:
\begin{equation}
\frac{du^{\alpha}}{ds}+\Gamma^{\alpha}_{\mu\nu}u^{\mu}u^{\nu}=\mathfrak{f}^{\alpha} ~,
\end{equation}
where $u^\mu$ denotes the particle's 4-velocity and the extra force, per unit mass, is given by:
\begin{equation}
\mathfrak{f}^{\alpha}=\frac{1}{\rho+p}\left[\frac{f'_2(R)}{f_2(R)}\left(\mathcal{L}_m-p\right)\nabla_{\nu}R-\nabla_{\nu}p\right]V^{\alpha\nu}~,
\end{equation}
where $V^{\alpha\nu} = g^{\alpha\nu} + u^{\alpha}u^{\nu}$ is the projection operator. We should also note the dependence on the matter Lagrangian density choice. This feature lifts the degeneracy that exists in GR whether $\mathcal{L}=-\rho$ or $\mathcal{L}=p$ \cite{lagrangian choices GR}, since it yields different results for the extra force (See Ref. \cite{lagrangian choices} for a thorough discussion).

%
%

\section{The Boltzmann equation in the Newtonian limit of the NMC}
\label{sec:weakfield}

In Ref. \cite{nmcboltzmann}, the Boltzmann equation for these theories was derived, where the main consequence was the appearance of a term related to the extra force:
\begin{equation}
p^{\mu}\frac{\partial f}{\partial x^{\mu}}-\left(\Gamma^{\sigma}_{\mu\nu}p^{\mu}p^{\nu}-m^2\mathfrak{f}^{\sigma}\right)\frac{\partial f}{\partial p^{\sigma}}=\left(\frac{\partial f}{\partial \tau^*}\right)_{coll.}~.
\end{equation}

Let us now study the case of dust, $p=0$, where the matter Lagrangian has a clear choice $\mathcal{L}=-\rho$. Furthermore we shall study the Newtonian level of the modified gravity model, Eq. (\ref{modelo}). We should note that the geodesic equation reads now \cite{martins}:
\begin{equation}
\frac{d^2x^i}{dt^2}=\partial_i\left[\frac{g_{tt}+1}{2}-\ln |f_2|\right] -\frac{\partial_i p}{\rho+p}~,
\end{equation}
from which one can define a NMC potential, $\Phi_c:=\ln|f_2|$ \cite{martins}. This implies that the nonminimally coupled Boltzmann equation reads in the absence of collisions and pressure gradients:
\begin{equation}
\frac{\partial f}{\partial t}+\vec{v}\cdot\nabla f - \nabla\left(\Phi+\Phi_c\right)\cdot\frac{\partial f}{\partial \vec{v}}=0~.
\label{eqn:weakboltzmann}
\end{equation}

Furthermore, the metric field equations of this model can be Taylor expanded considering corrections up to $c^{-2}$:
\begin{eqnarray}
R &&\sim R^{(2)}\equiv \delta R\\
f^n(R) &&\sim f^n(0)+f^{n-1}(0)R^{(2)}
\end{eqnarray}

Thus, at $\mathcal{O}(2)$, the 00-component of the field equations (\ref{fieldequations}) and the trace equation (\ref{trace}) become:
\begin{eqnarray}
&&f'_1(0)\delta G_{00}=\nabla^2\left[f''_1(0)\delta R+\frac{f'_2(0)}{\kappa}\delta\mathcal{L}_m\right]+\frac{f_2(0)}{2\kappa}\delta T_{00}^{(0)} ~,\\
&& f'_1(0)\delta R=3\nabla^2\left(f''_1(0)\delta R-\frac{f'_2(0)}{\kappa}\delta\mathcal{L}_m \right)+\frac{f_2(0)}{2\kappa}\delta T^{(0)}~,
\end{eqnarray}
where $f_1(0)=0$ because of the field equations at zeroth order. We point out that this expansion is performed around the Minskowski spacetime where at lowest order does not exist matter fields. However, at linear level the fluctuations of the components of the energy-momentum tensor correspond to matter fields, hence $\delta T_{\mu\nu}=\rho \delta^0_ {\mu}\delta^0_{\nu}$ and $\delta T=-\rho$. This situation contrasts with the study in the context of gravitational waves of Ref. \cite{nmcgw} where there still existed some residual background in the form of a cosmological constant or a dark energy-like fluid.

The obvious choice for the metric field is:
\begin{equation}
g_{\mu\nu}=diag(-1-2\Phi,1-2\Psi,1-2\Psi,1-2\Psi) ~,
\label{eqn:metric}
\end{equation}
where both $|\Phi|~,|\Psi| \ll 1$ and correspond to the Bardeen gauge invariant potentials. Thus, the 00-components of the Ricci tensor are $\delta R_{00}=\nabla^2\Phi$,  and the scalar curvature $\delta R=2\nabla^2\left(2\Psi-\Phi\right)$.

%

Inserting this metric in the perturbed metric field equations and their trace, we get:
\begin{eqnarray}
\left\{
\begin{array}{ll}
&2\alpha\nabla^4 (2\Psi-\Phi)-2\nabla^2\Psi = \beta\nabla^2\rho-\frac{\gamma}{2}\rho\\
&6\alpha\nabla^4(2\Psi-\Phi)-2\nabla^2(2\Psi-\Phi)=3\beta\nabla^2\rho-\frac{\gamma}{2}\rho ~,
\end{array}
\right.
\label{eqn:potentials0}
\end{eqnarray}
where $\alpha:=f''_1(0)/f'_1(0)$, $\beta:=f'_2(0)/\left(\kappa f'_1(0)\right)$, \break $\gamma:=f_2(0)/\left(\kappa F_1(0)\right)$, and the term $\nabla^2\rho$ comes directly from the non-minimal coupling. The usual Poisson equation in GR is retrieved by setting $\Phi=\Psi, ~f_2(R)=1$.

Fourier transforming the previous equations and adding both of them, one gets the following expression relating the two potentials:

\begin{equation}
\left(1+4\alpha k^2\right) \tilde{\Psi} = \left(1+2\alpha k^2\right)\tilde{\Phi}-\beta \tilde{\rho} ~,
\label{eqn:potentials1}
\end{equation}
where the tilde notation refers to the Fourier transform of the functions underneath it.

In fact, this equation deserves a few comments. If we have considered cosmological perturbations of the form of Eq. (\ref{eqn:metric}) in the metric field equations, we would have found that the ij-components yield a general condition relating both Bardeen invariant potentials \cite{frazao}:
\begin{equation}
\Phi-\Psi = -\delta \ln \left(f'_1(R)+f'_2(R)\mathcal{L}\right)~.
\end{equation}

Further expanding this condition, keeping terms up to $\mathcal{O}(1/c^2)$ and choosing $\mathcal{L}=-\rho$ we get the same relation, after a Fourier transform, as Eq. (\ref{eqn:potentials1}), where we have first expanded around a static Minkowsky background. The term $\beta \tilde{\mathcal{L}}=-\beta\tilde{\rho}$ appears due to the non-minimal coupling between matter and curvature. In the limit where $f_2(0)=1$, we retrieve the relation in $f(R)$ theories: $\tilde{\Psi}=\frac{1+2\alpha k^2}{1+4\alpha k^2}\tilde{\Phi}$. Moreover, when $f_1(R)=R \iff \alpha = 0$, we get $\tilde{\Psi}=\tilde{\Phi}$, which is the General Relativity's condition.

In fact, we can relate both Bardeen gauge invariant potentials without the need of the dependence on the matter Lagrangian choice by sorting out a different linear combination of both equations of the system of Eq. (\ref{eqn:potentials0}):
\begin{equation}
\tilde{\Psi}=\frac{\gamma+2\alpha\gamma k^2+ 2\beta k^2}{\gamma+4\alpha \gamma k^2-2\beta k^2}\tilde{\Phi}~.
\end{equation}

Now, the two potentials can be decoupled into two Poisson-like equations, resorting to the inverse Fourier transform for the real space, for each one:
\begin{eqnarray}
\left\{
\begin{array}{ll}
\left(3\alpha \nabla^4-\nabla^2\right)\Phi=\left(\alpha \gamma - \frac{\beta}{2}\right)\nabla^2 \rho-\frac{\gamma}{4}\rho \\
 \left(3 \alpha \nabla^4-\nabla^2\right)\Psi= \left(\frac{\alpha\gamma+\beta}{2}\right)\nabla^2 \rho -\frac{\gamma}{4}\rho ~.
\end{array}
\right.
\label{eqn:separatedpotentials}
\end{eqnarray}

These results are the basis for numerical solvers from nonlocal optics to simulate the dynamics of N-body systems in the non-minimal matter-curvature coupling model of Refs. \cite{numerico1,numerico2}, as they allow the study of weak field implications from modified gravity in the context of gravitational collapse. Furthermore we note that issues concerning stellar stability from modified Lane-Emden equation were addressed for the case of f(R) theories in Refs. \cite{lane1,lane2} and further generalised for the non-minimal matter-curvature model in Ref. \cite{lane3} together with the Tolman-Oppenheimer-Volkoff equation for a spherically symmetric body of isotropic material, where in both theories different solutions with respect to the standard theory were found. This equation aims at describing the inner structure of a thermodynamic self-gravitating system provided an polytropic fluid equation of state. A further stability criterion can be analysed in what concerns the  causes the collapse of interstellar gas clouds which lead to star formation. This is the so-called Jeans instability criterion. It was analysed in the context of f(R) theories in Refs. \cite{capozziello,vainio} and which shall be further analysed in the context of the non-minimal coupling alternative gravity model in the next section.


\section{Jeans instability}
\label{sec:jeans}

The equilibrium state, denoted with the subscript "0", is assumed to be homogeneous and time-independent. Let us consider a small departure from this equilibrium state \cite{binney}:
\begin{eqnarray}
&f(r, v, t) = f_0(r, v) +\epsilon f_1(r, v, t) ~,\\
&\Phi(r, t) = \Phi_0(r) + \epsilon \Phi_1(r, t) ~,\\
&\Psi(r, t) = \Psi_0(r) + \epsilon \Psi_1(r, t),
\end{eqnarray}
where $\epsilon \ll 1$.

Thus, we can set $f_0(x, v, t) =f_0(v)$, and consider the so-called Jeans "swindle" to set both equilibrium states $\Phi_0=0$ and $\Psi_0=0$. Hence the gravitational system will be characterised by the set of equations (\ref{eqn:weakboltzmann}) and (\ref{eqn:potentials0}) after a Fourier transform and taking into consideration that \break $\rho(\vec{x},t)=\int f(\vec{x},\vec{v},t)d\vec{v}$. Hence from the linearised weak field nonminimally coupled Boltzmann equation we get the following relation:

\begin{eqnarray}
&&-i\omega \tilde{f}_1+\vec{v}\cdot (i \vec{k} \tilde{f}_1)-i\vec{k}(\tilde{\Phi}_1+\tilde{\Phi}_{c1})\cdot \frac{\partial \tilde{f}_0}{\partial \vec{v}}=0 \iff \nonumber \\
&& \tilde{f}_1=\frac{\vec{k}\cdot \frac{\partial \tilde{f}_0}{\partial \vec{v}}}{\vec{v}\cdot \vec{k}-\omega} \left(\tilde{\Phi}_1+\tilde{\Phi}_{c1}\right)~.
\end{eqnarray}

We should now clarify the form of the perturbation $\tilde{\Phi}_{c1}$. Let us note that $\Phi_c=\ln f_2(R)$, from which follows that
\begin{equation}
\Phi_{c1}=\frac{f'_2(0)}{f_2(0)}\delta R = 2\epsilon\frac{\beta}{\gamma}\nabla^2(2\Psi_1-\Phi_1)~,
\end{equation}
which is a clearly subdominant term given the Jeans swindle. Hence, we can henceforth neglect it. We should note that if we aimed at studying the weak field implications of the non-minimal coupling effects, we would find that these are measurable and lead to important signatures, such as a stronger gravitational pull \cite{martins}, and shock waves in gravitational collapse \cite{numerico1,numerico2}.

Therefore, we can recast the first equality of Eq. (\ref{eqn:separatedpotentials}), in the Fourier space, together with the Jeans swindle:
\begin{equation}
1+\frac{(\alpha\gamma-\beta/2)k^2+\gamma/4}{k^2+3\alpha k^4}\int \frac{\vec{k}\cdot \frac{\partial f}{\partial \vec{v}}}{\vec{v}\cdot\vec{k}-\omega}d\vec{v}=0~.
\label{eqn:dispersion}
\end{equation}

This is the so-called dispersion relation. In the case of weak field of stellar systems, we can assume that the equilibrium distribution function follows a Maxwellian distribution function \cite{binney}:
\begin{equation}
f_0 = \frac{\rho_0}{(2\pi \sigma^2)^{3/2}}e^{-\frac{v^2}{2\sigma^2}}~,
\end{equation}
where $\rho_0$ is the characteristic density of each system and $\sigma$ is the standard deviation. We can impose that $\vec{k}=(k,0,0)$ without loss of generality. In the case of General Relativity, where $\alpha=\beta=0$ and $\gamma=1/\kappa$, the limit for instability is found by setting $\omega=0$, yielding the Jeans wavenumber \cite{binney}:
\begin{equation}
k_J^2:=k^2(\omega=0)=\frac{4\pi G \rho_0}{\sigma^2}~.
\end{equation}

This quantity defines the Jeans' length:
\begin{equation}
\lambda_J:=\frac{4\pi^2}{k_J^2}=\frac{\pi\sigma}{G\rho_0}~,
\end{equation}
which characterises the stability of perturbations on a sphere with the Jeans mass:
\begin{equation}
M_J := \frac{4\pi\rho_0}{3}\left(\frac{\lambda_J}{2}\right)^3=\frac{\pi}{6}\sqrt{\frac{1}{\rho_0}\left(\frac{\pi \sigma^2}{G}\right)^3}~.
\end{equation}

In fact, we can study unstable ($Re(\omega)=0 \wedge Im(\omega) > 0$), neutrally stable ($Re(\omega)\neq 0 \wedge Im(\omega) = 0$) and Landau damped solutions ($|Im(\omega)/Re(\omega)|\gtrsim \mathcal{O}(1)$) from the dispersion relation, Eq. (\ref{eqn:dispersion}). In this work, we are interested in the unstable modes. For the non-minimal matter-curvature coupling model, those are found by noting that the dispersion relation for the Maxwellian distribution can be recast as:
\begin{eqnarray}
&&\frac{k^2+3\alpha k^4}{(\alpha\gamma-\beta/2)k^2+ \gamma/4}=-\int_{-\infty}^{+\infty}\int_{-\infty}^{+\infty}\int_{-\infty}^{+\infty}\frac{k\frac{\partial f_0}{\partial v_x}}{v_x k - \omega}dv_xdv_ydv_z\nonumber \\
&& = D_1 \int_{-\infty}^{+\infty}\left[\int_{-\infty}^{+\infty}\left[\int_{-\infty}^{+\infty}\frac{k v_x e^{-\frac{v_x^2}{2\sigma^2}}}{k v_x - \omega}dv_x\right]e^{-\frac{v_y^2}{2\sigma^2}}dv_y\right]e^{-\frac{v_z^2}{2\sigma^2}}dv_z\nonumber\\
&& = \frac{\rho_0}{\sqrt{2\pi \sigma^2}\sigma^2}\int_{-\infty}^{+\infty}\frac{k v_x e^{-\frac{v_x^2}{2\sigma^2}}}{k v_x - \omega}dv_x\nonumber\\
&& = \frac{k_J^2}{4\pi G}\frac{1}{\sqrt{2\pi}}\int_{-\infty}^{+\infty} \frac{x e^{-\frac{x^2}{2}}}{x-B} dx ~,
\end{eqnarray}
where $D_1=\frac{\rho_0}{(2\pi \sigma^2)^{3/2}\sigma^2}$, and we have used the result for a Gaussian integral $\int_{-\infty}^{+\infty}e^{-\frac{y^2}{c^2}}dy= c\sqrt{\pi}$, together with the notation $x:=v_x/\sigma$ and $B := \omega/(k\sigma)$. By integrating by parts and considering purely imaginary frequencies $\omega = i w$, we find that the dispersion relation in these alternative theories of gravity read:
\begin{eqnarray}
&&\frac{k^2+3\alpha k^4}{(\alpha\gamma-\beta/2)k^2+ \gamma/4}=\nonumber\\
&&\frac{k_J^2}{4\pi G} \left[1+\frac{1}{\sqrt{2\pi}}e^{-\frac{B^2}{2}} B \left(-\pi \textrm{erfi}\left(\frac{B}{\sqrt{2}}\right) + \mathrm{Log}[-1]\right)\right]~,
\end{eqnarray}
where $\textrm{erfi}(x)=-i \textrm{erf}(i x)$ is the imaginary error function which is related to the error function $\textrm{erf}(x)=\frac{1}{\sqrt{x}}\int_{\infty}^{+\infty}e^{-t^2}dt$, and $\mathrm{Log}[x]$ is the analytic continuation of the logarithm function. Since, we are looking at unstable modes, we only aim at $\omega = i w$ solutions. Therefore, the previous result can be cast in the following form:
\begin{equation}
\frac{k^2+3\alpha k^4}{(\alpha\gamma-\beta/2)k^2+ \gamma/4}=\frac{k_J^2}{4\pi G} \left[1-\sqrt{\pi}e^{z^2} z \left(1-\textrm{erf}(z)\right)\right]~,
\end{equation}
where $z:=\frac{w}{\sqrt{2}k\sigma}$, and we used symmetry properties of the error function. This equation generalises the dispersion relation equation found in General Relativity \cite{binney} and in the context of $f(R)$ theories \cite{capozziello, vainio}. In fact, the right hand side of the previous equation decreases monotonically with respect to $z$, and the limit for instability occurs when $\omega=z=0$, i.e., $\textrm{erf}(0)=0$. Hence:
\begin{equation}
\frac{k^2+3\alpha k^4}{(\alpha\gamma-\beta/2)k^2+ \gamma/4}=\frac{k_J^2}{4\pi G} ~,
\label{eqn:ksolutions}
\end{equation}
which can be solved for $k^2$, yielding the following solutions:
\begin{equation}
k^2_{\pm}= \frac{-8\pi G + (2\alpha\gamma-\beta)k_J^2 \pm \sqrt{D_2}}{48\alpha \pi G}~,
\end{equation}
where the $\pm$ stands for each sign before the square root and $D_2=48 \alpha \gamma \pi G k_J^2   + (-8\pi G + (2\alpha\gamma-\beta)k_J^2)^2$. From this result we can write the modified Jeans' mass:
\begin{eqnarray}
&&\tilde{M}_J=\left(\frac{k_J^2}{k^2}\right)^{3/2}M_J=\nonumber\\
&&\left(\frac{48\alpha \pi G k^2_J}{-8\pi G + (2\alpha\gamma-\beta)k_J^2 \pm \sqrt{D_2}}\right)^{3/2}M_J~.
\end{eqnarray}

These results need to take into consideration further conditions which arise, for instance, from avoidance of other instabilities. In fact, we note that in order to find positive mass solutions, we need to require that  both numerator and denominator of the expression between parenthesis have the same sign. In addition, we need to require that the radicand (argument of the radical) is positively defined, i.e., \break $48 \alpha \gamma \pi G k_J^2   + (-8\pi G + (2\alpha\gamma-\beta)k_J^2)^2  > 0$.

Moreover, in order to keep gravitation attractive, a further condition appears from the metric field equations, \break $\frac{f_2(R)}{f'_1(R)-f'_2(R)\rho/\kappa}>0$ \cite{sequeira}, which is still valid at $R=0$, yielding:
\begin{equation}
\frac{\gamma}{1-\beta \rho}> 0~.
\end{equation}

A final remark arises from the avoidance of Dolgov-Kawasaki instabilities \cite{dolgov}, which in the case of the present modified gravity theory yields the condition \cite{sequeira}:
\begin{equation}
f_1''(R)+\frac{f_2''(R)\mathcal{L}}{\kappa}\geq 0~.
\end{equation}

Bearing these constraints in mind, we can now study the limits for Jeans' instability. To do so, let \break $\Delta_{\pm} := \frac{48\alpha \pi G k^2_J}{-8\pi G + (2\alpha\gamma-\beta)k_J^2 \pm \sqrt{48 \alpha \gamma \pi G k_J^2   + (-8\pi G + (2\alpha\gamma-\beta)k_J^2)^2}}$, where the subscripts are understood as the $+$ or $-$ solutions, and special cases can be readily scrutinised, such as f(R) theories or pure non-minimal matter-curvature coupling, and special cases as $2\alpha \gamma=\beta$.

\subsection{f(R) theories}

\begin{figure}[h!]
\centering
\includegraphics[scale=0.7]{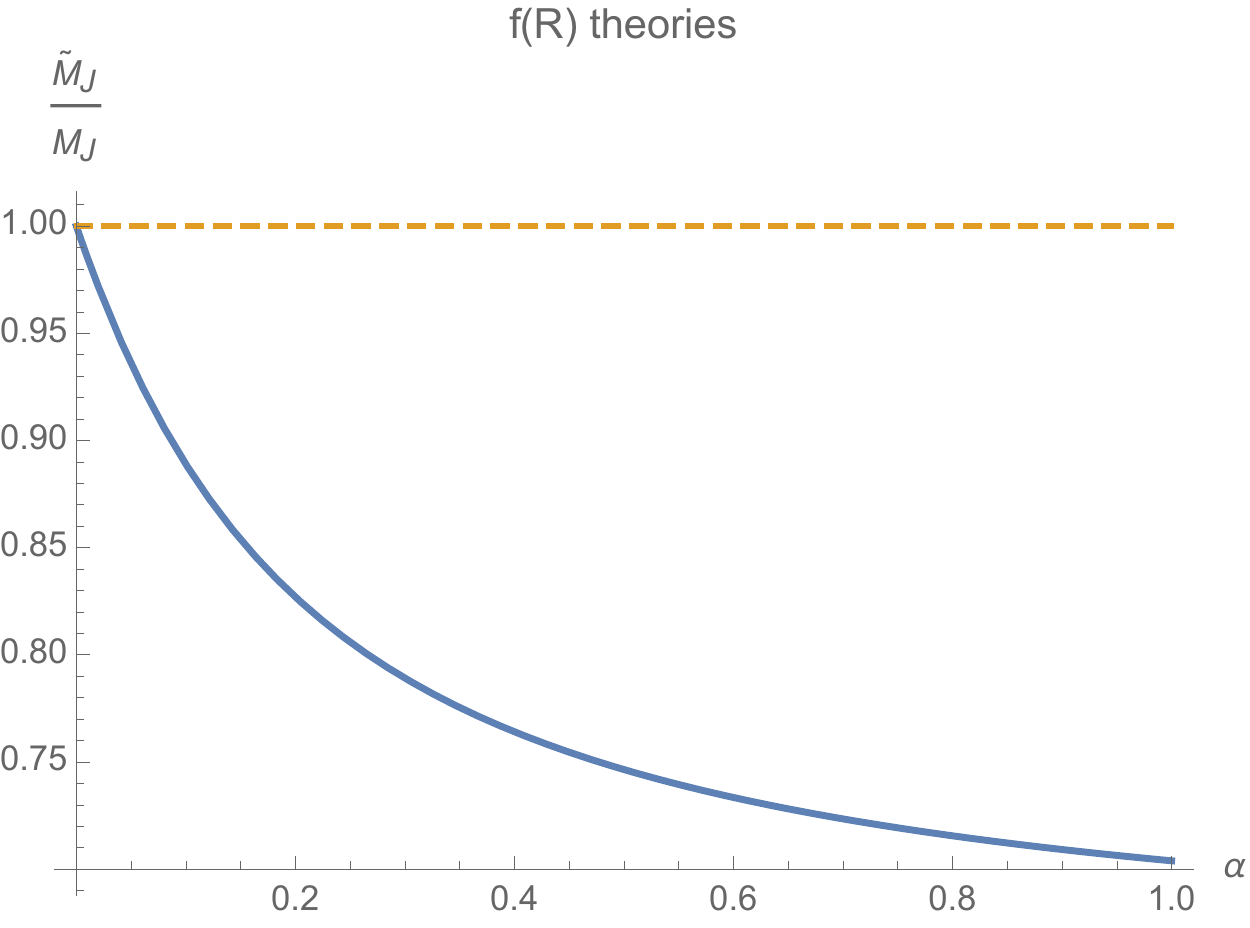}
\caption{Comparison between the GR (dashed line) and the f(R) theories (solid line) behaviours for the modified Jeans' mass.}
\label{fig:f(R)}
\end{figure}

In fact, f(R) theories can be found by setting $\alpha \neq 0, ~\beta=0, ~\gamma=1/\kappa$. This implies that we retrieve the results from Ref. \cite{vainio}:
\begin{equation}
\Delta_{\pm} = \frac{6\alpha k_J^2}{-1+4\alpha k^2_J \pm \sqrt{1+4\alpha k^2_J+16\alpha^2 k^4_J}}> 0 ~,
\end{equation} 
where bounds were found for the only viable solutions, $\Delta_+$, for the ratio $\tilde{M}_J/M_J\in (0.649519,1]$, given that for these models the Dolgov-Kawasaki instabilities are avoided provided $\alpha >0$. This behaviour is shown in Fig. (\ref{fig:f(R)}).

\subsection{Pure Non-Minimal Coupling}

On the other hand, we can set $f_1(R)=R$ but still consider that $f_2(R)\neq 1$. In order to tackle this situation, we need to solve Eq. (\ref{eqn:ksolutions}) once again but setting $\alpha=0$. Thus:
\begin{equation}
k^2_{\pm}=\frac{\gamma k_J^2}{16\pi G + 2 \beta k_J^2}~,
\end{equation}
which gives a new expression for $\Delta$:
\begin{equation}
\Delta=\frac{16\pi G + 2 \beta k_J^2}{\gamma} >0 ~.
\end{equation}

Therefore, either both $\beta, ~\gamma >0$ or $\gamma <0 \wedge \beta < -\frac{16\pi G}{2k_J^2}$. These are conditions for the non-minimal coupling function, $f_2(R)$, and its derivative evaluated at $R=0$. Therefore, only the case $\beta, ~\gamma >0$ is physically viable, as the remain would give repulsive gravity. In fact, $\gamma \in (0,1)$ represents the case when the corrected gravitational force is stronger than the Newtonian one and gives $\Delta \in (16\pi G, +\infty)$, whilst $\gamma \in (1,+\infty)$ denotes the opposite case when the correction leads to a weaker force than the Newtonian one and yields $\Delta \in (16\pi G, 2k_J^2)$ or $\Delta \in (2k_J^2,16\pi G)$ depending on the modulus of $k_J^2$. 

\subsection{$2\alpha \gamma - \beta=0$}

A third case is worth discussing, namely when $2\alpha \gamma = \beta$, where the expression for the $\Delta$ factor simplifies into the condition:
\begin{equation}
\Delta_{\pm} = \frac{6\alpha k^2_J}{-1 \pm \sqrt{\frac{3}{4\pi G} \alpha \gamma k_J^2   + 1}}>0 ~.
\end{equation}

In this case, we cannot exclude either $\Delta_+$ or $\Delta_-$ solutions since the condition from Dolgov-Kawasaki instability does not provide a way to assess whether solution is the most suitable one.

If $\alpha \to 0$ and $\gamma < +\infty$, then $\Delta_{\pm} \to 0$. However, if $\alpha \to +\infty$ and $\gamma < +\infty$, then $\Delta_{\pm} \to +\infty$. On the other side, if $\gamma \to 0$ and $\alpha < +\infty$, we obtain $\Delta_- \to -3\alpha k_J^2$ and $\alpha <0$, or $\Delta_+ \to +\infty$. The final case, if $\gamma \to +\infty$ and $\alpha < +\infty$, we obtain $\Delta_{\pm} \to 0$.

\section{Bok globules}
\label{sec:bok}

Bok globules are nearby isolated and simple-shaped clouds of interstellar gas and dust, with core temperatures of the order of $10~K$ and masses around $10~M_{\odot}$, and which can experience star formation \cite{Bok}. Furthermore, their masses are of the order of the corresponding Jeans' masses, which proves to an useful tool to distinguish between different models of gravity in what concerns to stability. Despite the fact that the formation process of Bok globules are not understood, and it is usual to assume homogeneous distribution and spherical symmetry, although they are ellipsoids \cite{ellipsoids}, for their cores in order to infer some properties, the available observational data on such gravitationally bound systems is sufficient for test the viability of the Jeans' criterion. This analysis was performed in Ref. \cite{vainio} in the context of f(R) theories resorting to data from Ref. \cite{Bok}. 

However, we should note that data from Ref. \cite{Bok} was found under some assumptions, namely the Bonnor-Ebert density profile, which is a solution of the Lane-Emdem equation provided the energy density at $r=0$ is nonsingular. Therefore, testing alternative theories of gravity resorting to this processed data may lack some features since, for instance, the estimated observational masses rely on the assumption that GR holds.

As shown in Ref. \cite{lane3}, in the perturbative regime of the non-minimal coupling model, assuming some power-law functions for the $f_1(R)$ and $f(R)$ functions, a non-trivial deviation from the standard Lane-Emdem equation is found, which leads to a dressed or effective mass. However, the choices for both functions of the model in \cite{lane3} are not the only viable ones, therefore a further discussion on those specific models would narrow down the observational outcomes of the model. Furthermore, the dressed masses may not be the physical or bare masses since the corrections are of gravitational nature, leading to a sum of gravitational and "real" components. Hence, we shall use the standard estimated observational masses as probes to test the viability and general physical consequences of the non-minimal matter-curvature model in what concerns the Jeans instability.

We further note that although the precision of the available data is not sufficient for more accurate conclusions, we can infer based on Table 1 from Ref. \cite{vainio} that a sufficient condition for finding the correct stability stage from Ref. \cite{Bok} seem to require that $\tilde{M}_J \lesssim (2/5) M_J$.

This phenomenological bound implies that for $f(R)$ theories, the limit for instability occurs when:
\begin{equation}
\alpha \sim \frac{15 (-5 + 20^{1/3})}{2 k_J^2 (-50 + 30\times 20^{1/3}  - 
   9\times  50^{1/3} + 50 k_J^4)}~,
\end{equation}
whilst the pure non-minimal coupling case yields:
\begin{equation}
\gamma \sim \left(16\pi G + 2 \beta k_J^2\right)\left(\frac{2}{5}\right)^{2/3}~.
\end{equation}

In its turn, the third case yields for both $\Delta_{\pm}$:
\begin{equation}
\gamma \sim 16 \pi G \left[ \left(\frac{2}{5}\right)^{2/3} + \frac{6}{5} \left(\frac{2}{5}\right)^{1/3} k_J^2 \alpha \right]~.
\end{equation}

These conditions for each of the three limits studied in this work seem to be physically viable and match the stable or unstable globules found in Ref. \cite{Bok}.

Moreover, in pure $f(R)$ theories, the modified Jeans' mass is lower than the standard one, which results in more efficient star formation processes and better agreement with data both from molecular clouds \cite{capozziello} and Bok globules \cite{vainio}.

In its turn, the non-minimal coupling model offers both possibilities: lower and higher modified Jeans masses in comparison with the ones from General Relativity. Since observational data favour models with lower Jeans masses in order to match observed stability conditions, this poses constraints on the functions of the non-minimal matter-curvature model and their derivatives. However, we should note that we can, in principle, find such regions even in cases other than the three special ones of the previous Section. Given this feature, even in cases where pure $f(R)$ fail to match data, this model seems to be more advantageous over the first. This should not be surprising as the model under study in the present work is a generalisation of the $f(R)$ theories.

\section{Conclusions}
\label{sec:conclusions}

In this work, we have analysed the weak field regime of the Boltzmann equation in the context of non-minimal matter-curvature coupling alternative gravity model. In this framework and resorting to the Jeans swindle, we found that the correction to the Newtonian potential is a subleading term, hence providing no effect on the linear expansion. However, the effects of the arbitrary functions of the scalar curvature are present in the form of three parameters which affect the relation between the two Bardeen gauge invariant potentials.

Applying the Jeans swindle, we generalised the results for the Jeans' criterion for gravitational instability, which led to a modified Jeans length and, consequently, a modified Jeans' mass. From these quantities we found some limits for instability in terms of the three parameters from the modification of gravity.

Bok globules are still not well understood gravitationally bound systems, with masses of the order of the Jeans masses. This is an excellent laboratory to test modified gravity models in what concerns the Jeans' criterion for instabilities, which result in star formation. Although, more accurate data is mandatory in the future, we used them to test the viability of our results in a similar way as Ref. \cite{vainio}. By comparing the analysis from \cite{vainio} with data from Ref. \cite{Bok}, we found that a sufficient condition for viable models is to require that $\tilde{M}_J \sim (2/5) M_J$. This can, in principle, be easily found in the present model.

\end{document}